\documentclass[runningheads]{llncs}
\usepackage[T1]{fontenc}
\usepackage{graphicx}
\usepackage{amssymb,amsmath}

\DeclareMathOperator*{\argmin}{arg\,min}
\DeclareMathOperator{\Tr}{Tr}
\usepackage{xcolor}

\usepackage{hyperref,graphicx}
\usepackage{url}
\usepackage{textcomp}
\begin{document}
\title{mSPD-NN: A Geometrically Aware Neural Framework for Biomarker Discovery from Functional Connectomics Manifolds}
\titlerunning{A Geometrically Aware Neural Framework}
%
\author{Niharika S. D'Souza \inst{1} 
 \textsuperscript{*}, Archana Venkataraman \inst{2}}
\authorrunning{N.S. D'Souza et al.}
%
\institute{IBM Research, Almaden, San Jose, USA  \\ \email{Niharika.Dsouza@ibm.com} \and 
Dept. of Electrical and Computer Eng., Johns Hopkins University, Baltimore, USA}
\maketitle              
\begin{abstract}
Connectomics has emerged as a powerful tool in neuroimaging and has spurred recent advancements in statistical and machine learning methods for connectivity data. 
Despite connectomes inhabiting a matrix manifold, most analytical frameworks ignore the underlying data geometry. 
This is largely because simple operations, such as mean estimation, do not have easily computable closed-form solutions.
We propose a geometrically aware neural framework for connectomes, i.e., the mSPD-NN, designed to estimate the geodesic mean of a collections of symmetric positive definite (SPD) matrices.
The mSPD-NN is comprised of bilinear fully connected layers with tied weights and utilizes a novel loss function to optimize the matrix-normal equation arising from Fr\'echet mean estimation.
Via experiments on synthetic data, we demonstrate the efficacy of our mSPD-NN against common alternatives for SPD mean estimation, providing competitive performance in terms of scalability and robustness to noise.
We illustrate the real-world flexibility of the mSPD-NN in multiple experiments on rs-fMRI data and demonstrate that it uncovers stable biomarkers associated with subtle network differences among patients with ADHD-ASD comorbidities and healthy controls.
\end{abstract}

\section{Introduction}
Resting state functional MRI (rs-fMRI) measures steady state patterns of co-activation~\cite{fox2007spontaneous} (i.e., \textit{connectivity}) as a proxy for communication between brain regions. The `connectome' is a whole-brain map of these connections, often represented as a correlation or covariance matrix~\cite{lindquist2008statistical} or a network-theoretic object such as adjacency matrix or graph kernel~\cite{fornito2013graph}.
The rise of connectomics has spurred many analytical frameworks for group-wise diagnostics and biomarker discovery from this data. Early examples include statistical comparisons of connectivity features~\cite{lindquist2008statistical}, aggregate network theoretic measures~\cite{fornito2013graph}, and dimensionality reduction techniques~\cite{khosla2019machine,d2020joint}. More recently, the field has embraced deep neural networks to learn complex feature representations from both the connectome and the original rs-fMRI time series~\cite{bessadok2022graph,nandakumar2020multi,d2019integrating}. While these approaches have yielded valuable insights, they largely ignore the underlying geometry of the connectivity data. Namely, under a geometric lens, connectomes derived from rs-fMRI data lie on the manifold of symmetric positive definite (SPD) matrices. A major computational bottleneck for developing geometrically-aware generalizations~\cite{nguyen2019neural,banerjee2015nonlinear} is the estimation of the geodesic mean on SPD manifolds. This is a far more challenging problem than statistical estimation in Euclidean data spaces because extensions of elementary operations such as addition, subtraction, and distances on the SPD manifold entail significant computational overhead~\cite{moakher2005differential}. 

The most common approach for estimating the geodesic mean on the SPD manifold is via gradient descent~\cite{pennec2006riemannian}. While this method is computationally efficient, it is highly sensitive to the step size. To mitigate this issue, Riemannian optimization methods~\cite{jeuris2015riemannian}, the majorization-maximization algorithm~\cite{zhang2013majorization}, and fixed-point iterations~\cite{congedo2017fixed} can be used. While these extensions have desirable convergence properties, this comes at the cost of increased computational complexity, meaning they do not scale well to higher input dimensionality and larger numbers of samples~\cite{congedo2015approximate}. In contrast, the work of~\cite{congedo2015approximate} leverages the approximate joint diagonalization~\cite{pham2001joint} of matrices on the SPD manifold. While this approach provides guaranteed convergence to a fixed point, the accuracy and stability of the optimization is sensitive to the deviation of the data from the assumed common principal component (CPC) generating process. Taken together, existing methods for geodesic mean estimation on the SPD manifold poorly balance accuracy, robustness and computational complexity, which makes them difficult to fold into a larger analytical framework for connectomics data.

We propose a novel end-to-end framework to estimate the geodesic mean of data on the SPD manifold. Our method, the Geometric Neural Network (mSPD-NN), leverages a matrix autoencoder formulation~\cite{d2021matrix} that performs a series of bi-linear transformations on the input SPD matrices. This strategy ensures that the estimated  mean remains on the manifold at each iteration. Our loss function for training 
approximates the first order matrix-normal condition arising from Fr\'echet mean estimation~\cite{moakher2005differential}. Using conventional backpropagation via stochastic 
optimization, the mSPD-NN automatically learns to estimate the geodesic mean of the input data. 
We demonstrate the robustness of our framework using simulation studies and show that mSPD-NN can handle input noise and 
high-dimensional data. Finally, we use the mSPD-NN for various groupwise discrimination tasks (feature selection, classification, clustering) on functional connectivity data and discover consistent 
biomarkers that distinguish between patients diagnosed with ADHD-Autism comorbidities and healthy controls. 

\section{Biomarker Discovery from Functional Connectomics Manifolds via the mSPD-NN}

Let matrices $\{\mathbf{\Gamma}_{n}\}^{N}_{n=1} \in \mathcal{M}$ be a collection of $N$ functional connectomes belonging to the manifold $\mathcal{M}$ of Symmetric Positive Definite (SPD) matrices of dimensionality $P \times P$,  i.e. $\mathcal{M} \in \mathcal{P}^{+}_{P}$ (and a real and smooth Reimannian manifold). We can define an inner product that varies smoothly at each vector $\mathcal{T}_{\mathbf{\Gamma}}(\mathcal{M})$ in the tangent space defined at any point $\mathbf{\Gamma} \in \mathcal{M}$. Finally, a \textit{geodesic} denotes the shortest path joining any two points on the manifold along the manifold surface.

\medskip
\noindent\textbf{Geodesic Mappings:} The matrix exponential and the matrix logarithm maps allow us to translate geodesics on the manifold back and forth to the local tangent space at a reference point. The matrix exponential mapping translates a vector $\mathbf{V} \in \mathcal{T}_{\mathbf{\Phi}}(\mathcal{M})$ in the tangent space at $\mathbf{\Phi} \in \mathcal{M}$ to a point on the manifold $\mathbf{\Gamma} \in \mathcal{M}$ via the geodesic emanating from $\mathbf{\Phi}$. Conversely, the matrix logarithm map translates the geodesic between $\mathbf{\Phi} \in \mathcal{M}$ to $\mathbf{\Gamma} \in \mathcal{M}$ back to the tangent vector  $\mathbf{V} \in \mathcal{T}_{\mathbf{\Phi}}(\mathcal{M})$. Mathematically, these operations are parameterized as:
\begin{gather}
    \mathbf{\Gamma} = \textbf{Expm}_{\mathbf{\Phi}}(\mathbf{V}) = \mathbf{\Phi}^{1/2}\textbf{expm}(\mathbf{\Phi}^{-1/2}\mathbf{V}\mathbf{\Phi}^{-1/2})\mathbf{\Phi}^{1/2} \\
    \mathbf{V} = \textbf{Logm}_{\mathbf{\Phi}}(\mathbf{\Gamma}) = \mathbf{\Phi}^{1/2}\textbf{logm}(\mathbf{\Phi}^{-1/2}\mathbf{\Gamma}\mathbf{\Phi}^{-1/2})\mathbf{\Phi}^{1/2}
\end{gather}
Here, $\textbf{expm}(\cdot)$ and $\textbf{logm}(\cdot)$ refer to the matrix exponential and logarithm respectively, each requiring an eigenvalue decomposition of the argument matrix, a point-wise transformation of the eigenvalues, and a matrix reconstruction.

\medskip
\noindent\textbf{Distance Metric:} Given two connectomes $\mathbf{\Gamma}_{1},\mathbf{\Gamma}_{2} \in \mathcal{M}$, the Fisher Information distance between them is the length of the geodesic connecting the two points:
\begin{equation}
\delta_{R}(\mathbf{\Gamma}_{1},\mathbf{\Gamma}_{2}) =  {\vert\vert{\textbf{logm}(\mathbf{\Gamma}^{-1}_{1} \mathbf{\Gamma}_{2})}\vert\vert}_{F} =  {\vert\vert{\textbf{logm}(\mathbf{\Gamma}^{-1}_{2} \mathbf{\Gamma}_{1})}\vert\vert}_{F}, 
\label{dist}
\end{equation}
where ${\vert\vert{\cdot}\vert\vert}_{F}$ denotes the Frobenius norm. The Reimannian norm of $\mathbf{\Gamma}$ is the geodesic distance from the identity matrix $\mathcal{I}$ i.e. ${\vert\vert{\mathbf{\Gamma}}\vert\vert}_{R} = {\vert\vert{\textbf{logm}(\mathbf{\Gamma})}\vert\vert}_{F}  $

\begin{figure}[t]
    \centering
    {\includegraphics[scale=0.40]{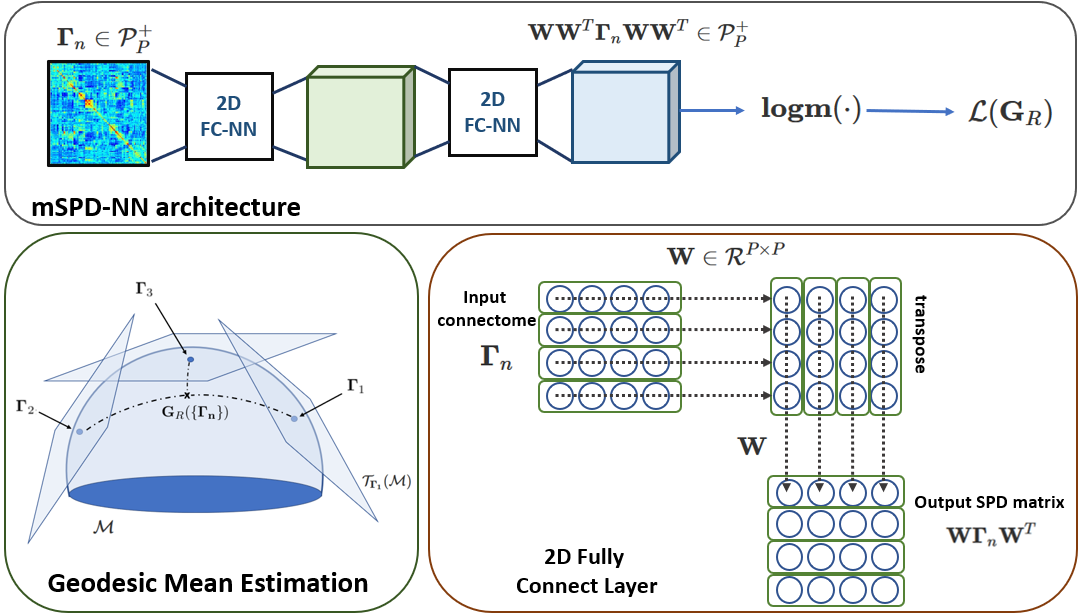}}
    \caption{\textbf{The mSPD-NN architecture:} The input is transformed by a cascade of 2D fully connected layers. The matrix logarithm function is used to obtain the matrix normal form, which serves as the loss function for mSPD-NN during training.}
    \label{fig:GeoNN}
\end{figure}

\subsection{Geodesic Mean Estimation via the mSPD-NN:} 
The geodesic mean of $\{\mathbf{\Gamma}_{n}\}$ is defined as the matrix $\mathbf{G}_{R} \in \mathcal{M}$ whose sum of squared geodesic distances (Eq.~(\ref{dist})) to each element is minimal~\cite{moakher2005differential}.
\begin{equation}
    \mathcal{G}_{R}(\{\mathbf{\Gamma}_{n}\}) = \argmin_{\mathbf{G}_{R}}  \mathbf{L}(\mathbf{G}_{R}) = \argmin_{\mathbf{G}_{R}} \sum_{n}{\vert\vert{\textbf{logm}(\mathbf{G}_{R}^{-1}\mathbf{\Gamma}_{n})}\vert\vert}^{2}_{F}
    \label{Geomean}
\end{equation}
A pictorial illustration is provided in the green box in Fig~\ref{fig:GeoNN}. While Eq.~(\ref{Geomean}) does not have a closed-form solution for $N>2$, it is also is convex and smooth with respect to the unknown quantity $\mathbf{G}_{R}(\cdot)$~\cite{moakher2005differential}. To estimate population means from the connectomes, mSPD-NN makes use of Proposition 3.4 from~\cite{moakher2005differential}.

\medskip
{\noindent\textbf{Proposition 1:} The geodesic mean $\mathbf{G}_{R}$ of a collection of $N$ SPD matrices $\{\mathbf{\Gamma}_{n}\}$ is the unique symmetric positive-definite solution to the nonlinear matrix equation $\sum_{n} \textbf{logm} (\mathbf{G}^{-1/2}_{R}\mathbf{\Gamma}_{n}\mathbf{G}^{-1/2}_{R})    =\mathbf{0}$. $\mathbf{0}$ is a $P \times P$ matrix of all zeros.}

\medskip
\noindent\textbf{\textit{Proof:}}
The proof follows by computing the first order necessary (and here, sufficient) condition for optimality for Eq.~(\ref{Geomean}). First, we express the derivative of a real-valued function of the form $\mathbf{H}(\mathbf{S}(t)) = 
\frac{1}{2}{\vert\vert{ \textbf{logm} (\mathbf{C}^{-1} \mathbf{S}(t)) }\vert\vert}_{F}^2$ with respect to $t$. In this expression, the argument $ \mathbf{S}(t) = \mathbf{\mathbf{G}_{R}}^{1/2}\textbf{expm}(t\mathbf{A})\mathbf{G_{R}}^{1/2}$ is the geodesic arising
from $\mathbf{G}_{R}$ in the direction of $\mathbf{\Delta} = \dot{\mathbf{S}}(\mathbf{0})= \mathbf{\mathbf{G}_{R}}^{1/2}\mathbf{A}\mathbf{G_{R}}^{1/2}$, and the matrix $\mathbf{C} \in \mathcal{P}^{+}_{P}$ is a constant SPD matrix of dimension $P$.  By using the cyclic properties of the trace function and the distributive equivalence of $\textbf{logm}(\mathbf{A}^{-1}[\mathbf{B}]\mathbf{A}) = \mathbf{A}^{-1}[\textbf{logm}(\mathbf{B})]\mathbf{A}$, we obtain the following condition:
\begin{gather*}
\mathbf{H}(\mathbf{S}(t)) = \frac{1}{2}{{\vert\vert{\textbf{logm}(\mathbf{C}^{-1/2} \mathbf{S}(t)\mathbf{C}^{-1/2})}\vert\vert}_F^2 } 
\end{gather*}
By the symmetry of the term \ \ $\textbf{logm}(\mathbf{C}^{-1/2} \mathbf{S}(t)\mathbf{C}^{-1/2})$ \ \ {we have that:} 
\begin{gather*}
 \therefore {\frac{d}{dt}{\mathbf{H}(\mathbf{S}(t))}}\Bigr|_{\substack{t=0}} = \frac{1}{2} \frac{d}{dt}{\Tr\Big([\textbf{logm}(\mathbf{C}^{-1/2} \mathbf{S}(t) \mathbf{C}^{-1/2})]^2\Big)} \Bigr|_{\substack{t=0}} \\  \therefore {\frac{d}{dt}{\mathbf{H}(\mathbf{S}(t))}}\Bigr|_{\substack{t=0}} = \Tr\Big([\textbf{logm}(\mathbf{C}^{-1}\mathbf{G}_{R}) \mathbf{G}_{R}^{-1}\mathbf{\Delta}]\Big) = \Tr[\mathbf{\Delta}\textbf{logm}(\mathbf{C}^{-1}\mathbf{G}_{R}) \mathbf{G}_{R}^{-1}] \\
 \therefore \nabla{\mathbf{H}} = \textbf{logm}(\mathbf{C}^{-1}\mathbf{G}_R)\mathbf{G}^{-1}_{R} = \mathbf{G}^{-1}_{R}  \textbf{logm}(\mathbf{G}_R\mathbf{C}^{-1})
\end{gather*}
Notice that since $\nabla{\mathbf{H}}$ is symmetric, it belongs to the tangent space $\mathcal{S}_{P}$ of $\mathcal{P}^{+}_{P}$. Therefore, we express the gradient of $\mathbf{L}(\mathbf{G}_{R})$ defined in Eq.~(\ref{Geomean}), as follows:
\begin{gather*}
\mathbf{L}(\mathbf{G}_{R}) = \sum_{n} {\vert\vert{\textbf{logm}(\mathbf{G}_{R}^{-1}\mathbf{\Gamma}_{n})}\vert\vert}^{2}_{F}  \ \ \ 
\implies \nabla \mathbf{L}(\mathbf{G}_{R}) = \mathbf{G}^{-1}_{R} \sum_{n}\textbf{logm}({\mathbf{G}_{R}\mathbf{\Gamma}^{-1}_{n}}) \\
    \therefore
\argmin_{\mathbf{G}_{R}}\mathbf{L}(\mathbf{G}_{R}) \implies  \sum_{n}\textbf{logm}({\mathbf{G}_{R}\mathbf{\Gamma}^{-1}_{n}})  = \sum_{n}\textbf{logm}({\mathbf{G}^{-1/2}_{R}\mathbf{\Gamma}_{n}\mathbf{G}^{-1/2}_{R}}) = \mathbf{0}
\end{gather*}
The final step uses the property that $\mathbf{L}(\mathbf{G}_{R})$ is a sum of convex functions, with the first order stationary point is the necessary and sufficient condition being the unique minima. \textit{Denoting $\mathbf{G}^{-1/2}_{R} = \mathbf{V} \in \mathcal{P}_{P}^{+}$, the matrix multiplications in the argument of the $\textbf{logm}(\cdot)$ term can be efficiently expressed within the feed-forward operations of a neural network with unknown parameters~$\mathbf{V}$.} 

\subsection{mSPD-NN Architecture} The mSPD-NN uses the form above to perform geodesic mean estimation. The architecture is
illustrated in Fig.~\ref{fig:GeoNN}. The encoder of the mSPD-NN is a 2D fully-connected neural network (FC-NN)~\cite{dong2017deep} layer $\mathbf{\Psi}_{\text{enc}}(\cdot) : \mathcal{P}_{P}^{+} \rightarrow \mathcal{P}_{P}^{+} $ that projects the input matrices $\mathbf{\Gamma}_{n}$ into a latent representation. This mapping is 
computed as a cascade of two linear layers with tied weights $\mathbf{W} \in \mathcal{R}^{P \times P}$, i.e., $\mathbf{\Psi}_{\text{enc}}(\mathbf{\Gamma}_{n}) = \mathbf{W}\mathbf{\Gamma}_{n}\mathbf{W}^{T}$ The decoder $\mathbf{\Psi}_{dec}(\cdot)$ has the same architecture as the encoder, but with transposed weights $\mathbf{W}^{T}$. The overall transformation can be written as:
\begin{equation}
\text{mSPD-NN}(\mathbf{\Gamma}_{n})   = \mathbf{\Psi}_{\text{dec}}(\mathbf{\Psi}_{\text{enc}}(\mathbf{\Gamma}_{n})) = \mathbf{W}\mathbf{W}^{T}(\mathbf{\Gamma}_{n})\mathbf{W}\mathbf{W}^{T} = \mathbf{V}(\mathbf{\Gamma}_{n})\mathbf{V}
\end{equation}
where $\mathbf{V} \in \mathcal{R}^{P\times P}$ and is symmetric and positive definite by construction. We would like our loss function to minimize Eq.~(\ref{Geomean}) in order to estimate the first order stationary point as $\mathbf{V} =\mathbf{G}^{-1/2}_{R}$, 
and therefore devise the following loss:
\begin{equation}
    \mathcal{L} (\cdot) = \frac{1}{P^2} {\Big\vert\Big\vert{\frac{1}{N}\sum_{n}\textbf{logm}\Big[{\mathbf{W}\mathbf{W}^{T}(\mathbf{\Gamma}_{n})\mathbf{W}\mathbf{W}^{T}}}\Big]\Big\vert\Big\vert}_{F}^{2}
    \label{loss}
\end{equation}
Formally, an error of $\mathcal{L}(\cdot) = 0 $ implies that the argument satisfies the matrix normal equation exactly under the parameterization $\mathbf{V} =\mathbf{W}\mathbf{W}^{T} = \mathbf{G}^{-1/2}_{R}$. Therefore, Eq.~(\ref{loss}) allows us to estimate the geodesic mean on the SPD manifold. We utilize standard backpropagation to optimize Eq.~(\ref{loss}). From an efficiency standpoint, the mSPD-NN architecture maps onto a relatively shallow neural network. Therefore, this module can be easily integrated into other deep learning inference frameworks for example, for batch normalization on the SPD manifold. This flexibility is the key advantage over classical methods, in which integrating the geodesic mean estimation within a larger framework is not straightforward. Finally, the extension of Eq.~(\ref{loss}) to the estimation of a weighted mean (with positive weights $\{w_{n}\}$) also follows naturally as a multiplier in the summation.

\medskip
\noindent\textbf{Implementation Details:} We train mSPD-NN for a maximum of $100$ epochs with an initial learning rate of $0.001$ decayed by $0.8$ every $50$ epochs. The tolerance criteria for the training loss is set at $1e^{-4}$. mSPD-NN implemented in PyTorch (v1.5.1), Python 3.5 and experiments were run on an 4.9 GB Nvidia K80 GPU. We utilize the ADAM optimizer during training and a default PyTorch initialization for the model weights. To ensure that $\mathbf{W}$ is full rank, we add a small bias to the weights, i.e., $\tilde{\mathbf{W}} = \mathbf{W} + \lambda \mathcal{I}_{P}$ for regularization and stability.

\section{Evaluation and Results}

\subsection{Experiments on Synthetic Data}
We evaluate the scalability, robustness, and fidelity of mSPD-NN using simulated data. We compare the mSPD-NN against two popular mean estimation algorithms, the first being the Riemannian gradient descent~\cite{pennec2006riemannian} on the objective in Eq.~(\ref{Geomean}) and the second being the \textbf{A}pproximate Joint Diagonalization \textbf{L}og \textbf{E}uclidean (ALE) mean estimation~\cite{congedo2015approximate}, which directly leverages properties of the common principal components (CPC) data generating process~\cite{pham2001joint}.

Our synthetic experiments are built off the CPC model~\cite{jolliffe2016principal}. In this case, each input connectome $\mathbf{\Gamma}_{n} \in \mathcal{R}^{P \times P}$ is derived from a set of components $\mathbf{B} \in \mathcal{R}^{P \times P}$ common to the collection and a set of example specific (and strictly positive) weights across the components $\mathbf{c}_{n} \in \mathcal{R}^{(+) P \times 1}$. Let the diagonal matrix $\mathbf{C}_{n}$ be defined as $\mathbf{C}_{n} = \mathbf{diag}(\mathbf{c}_{n}) \in \mathcal{R}^{(+) P \times P}$. From here, we have $\mathbf{\Gamma}_{n} = \mathbf{B} \mathbf{C}_{n} \mathbf{B}^{T}$.

\medskip
\noindent\textbf{Evaluating Scalability:}
\label{noiseless}
 In the absence of corrupting noise, the theoretically optimal geodesic mean of the examples $\{\mathbf{\Gamma}_{n}\}_{n=1}^N$ can be computed as:
$\mathbf{G}^{*}_{R} = \mathbf{B} \ \mathbf{expm}\left[\frac{1}{N}\sum_{n=1}^N \mathbf{logm}(\mathbf{B}^{-1}\mathbf{\Gamma}_{n}\mathbf{B}^{-T})\right] \ \mathbf{B}^{T}$~\cite{congedo2015approximate}.
We evaluate the scalability of each algorithm with respect to the dataset dimensionality $P$ and the number of examples $N$ by comparing its output to this theoretical optimum.

We randomly sample columns of the component matrix~$\mathbf{B}$ from a standard normal, i.e., $\mathbf{B}[:,j] \sim \mathcal{N}(\mathbf{0},\mathcal{I}_{P}) \ \ \forall  \ \ j \in \{1,\dots,P\}$, where $\mathcal{I}_{P}$ is an identity matrix of dimension $P$. In parallel, we sample the component weights~$\mathbf{c}_{nk}$ according to $\mathbf{c}_{nk}^{1/2} \sim \mathcal{N}(0,1) \ \  \forall \ \ k \in \{1,\dots,P\}$.
To avoid degenerate behavior when the inputs are not full-rank, we clip $\mathbf{c}_{nk}$ to a minimum value of $0.001$. We consider two experimental scenarios. In \textbf{Experiment 1}, we fix the data dimensionality at $P=30$ and sweep the dataset size as $N \in \{5,10,20,50,100,200\}$. In \textbf{Experiment 2}, we fix the dataset size at $N=20$ and sweep the dimensionality as $P \in \{5,10,20,50,100,200\}$. For each parameter setting, we run all three estimation algorithms ten times using different random initializations.

We score performance based on the correctness of the solution and the execution time in seconds. Correctness is measured in two ways. First is the final condition fit $\mathcal{L}(\mathbf{G}^{\text{est}}_{R})$ from Eq.~(\ref{loss}), which quantifies the deviation of the solution from the first order stationary condition (i.e., $\mathcal{L}(\mathbf{G}^{\text{est}}_{R})=0$). Second is the normalized squared Riemannian distance $d_{\text{mean}} = {d^{2}_{R}(\mathbf{G}^{\text{est}}_{R},\mathbf{G}^{*}_{R}})/{\vert\vert{\mathbf{G}^{*}_{R}}\vert\vert}^2_{R}$ between the solution and the theoretically optimal mean. Lower values of the condition fit $\mathcal{L}(\mathbf{G}_{R})$ and deviation $d_{\text{mean}}$ imply a better quality solution. 

\begin{figure}[t]
    \centering
    \includegraphics[width=\textwidth]{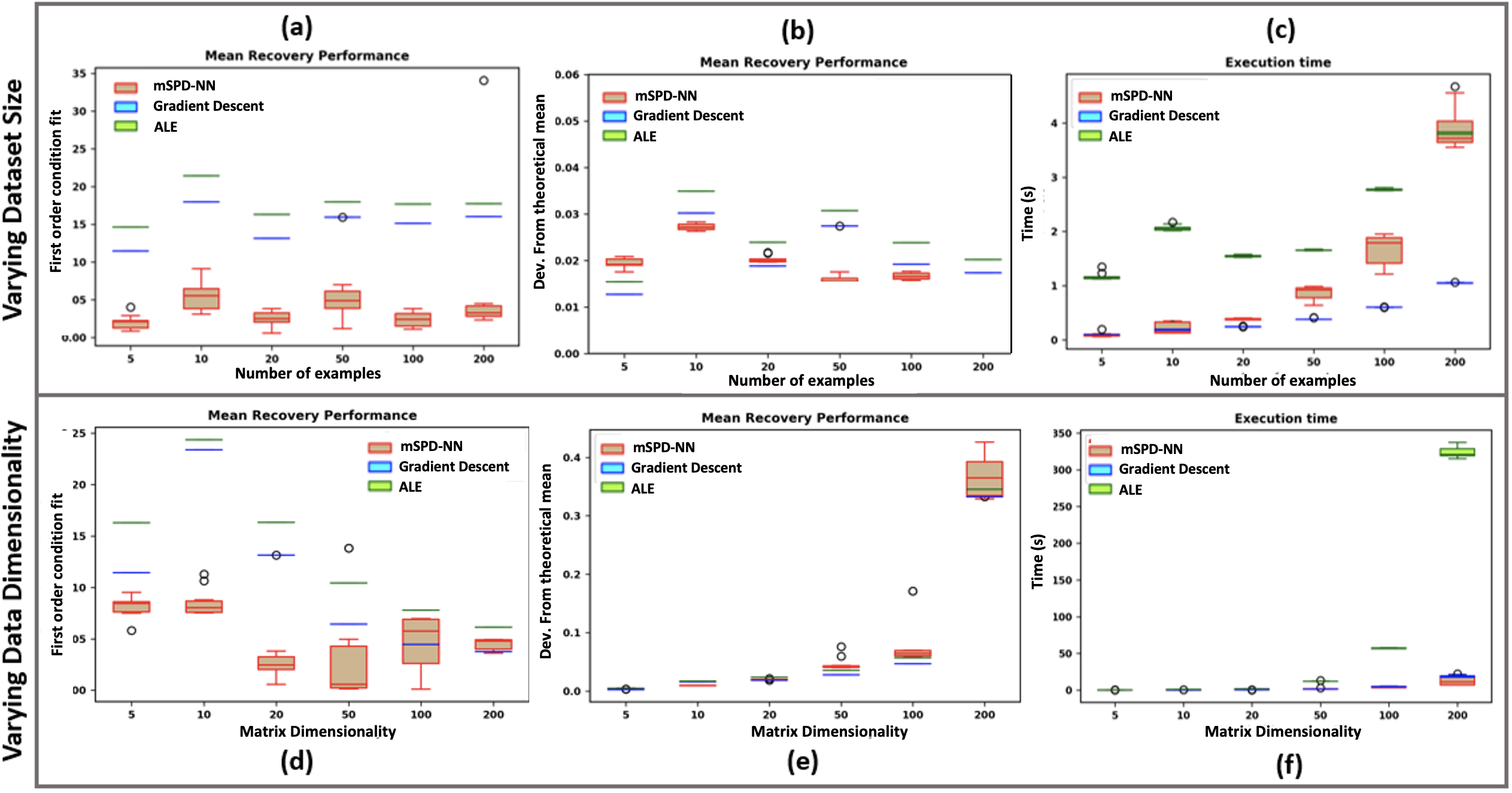}
    \caption{Evaluating the estimates from  mSPD-NN, gradient descent and ALE according to \textbf{(a) \textit{and} (d)} first-order condition fit (Eq.~\ref{loss}) \textbf{(b) \textit{and} (e)} deviation from the theoretical solution \textbf{(c) \textit{and} (f)} execution time for \textbf{varying dataset size~$N$ \textit{and} data dimension~$P$} respectively}
    \label{fig:FO_cond_dd}
\end{figure}

Fig.~\ref{fig:FO_cond_dd} illustrates the performances of mSPD-NN, gradient descent and ALE mean estimation algorithms. Figs.~\ref{fig:FO_cond_dd}(a)~and~(d) plot the first-order condition fit $\mathcal{L}(\mathbf{G}^{\text{est}}_{R})$ when varying the dataset size~$N$ (Experiment 1) and the matrix dimensionality~$P$ (Experiment 2), respectively. Likewise, Figs.~\ref{fig:FO_cond_dd}(b)~and~(e) plot the recovery performance for each experiment. We observe that the first order condition fit for the mSPD-NN is better than the ALE for all settings, and better than the gradient descent for most settings. We note that the recovery performance of mSPD-NN is better than the baselines in most cases while being a close approximation in the remaining ones. Finally, Figs.~\ref{fig:FO_cond_dd}(c)~and~(f) illustrate the time to convergence for each algorithm. As seen, the performance of mSPD-NN scales with dataset size but is nearly constant with respect to  dimensionality. In all cases, it either beats or is competitive with ALE.

\medskip
\noindent\textbf{Robustness to Noise:}
\label{robustness}
Going one step further, we evaluate the efficacy of the mSPD-NN framework when there is deviation from the ideal CPC generating process. In this case, we add rank-one structured noise to obtain the input data: $\mathbf{\Gamma}_{n} = \mathbf{B} \mathbf{C}_{n} \mathbf{B}^{T} + \frac{1}{P}\mathbf{x}_n\mathbf{x}_n^{T}$. As before, the bases and coefficients are randomly sampled as $\mathbf{B}[:,j] \sim \mathcal{N}(\mathbf{0},\mathcal{I}_{P})$ and  $\mathbf{c}_{nj}^{1/2} \sim \mathcal{N}(0,1) \ \  \forall \ \ j \in \{1,\dots,P\}$. In a similar vein, the structured noise is generated as $\mathbf{x}_n \sim \mathcal{N}(\mathbf{0},\sigma^2 \mathcal{I}_{P}) \in \mathcal{R}^{P \times 1}$, with $\sigma^2$ controlling the extent of the deviation. For this experiment, we set $P=30, N=20$ and vary the noise over the range $[0.2-1]$ in increments of $0.1$. One caveat in this setup is that the theoretically optimal mean defined previously and cannot be used to evaluate performance. Hence, we report only the first-order condition fit $\mathcal{L}(\mathbf{G}_{R})$ We also calculate the pairwise concordance $d_{\text{weights}}$ of the final mSPD-NN weights for different initializations. 
\begin{figure}[t!]
    \centering
   \fbox{ \includegraphics[width=0.67\textwidth]{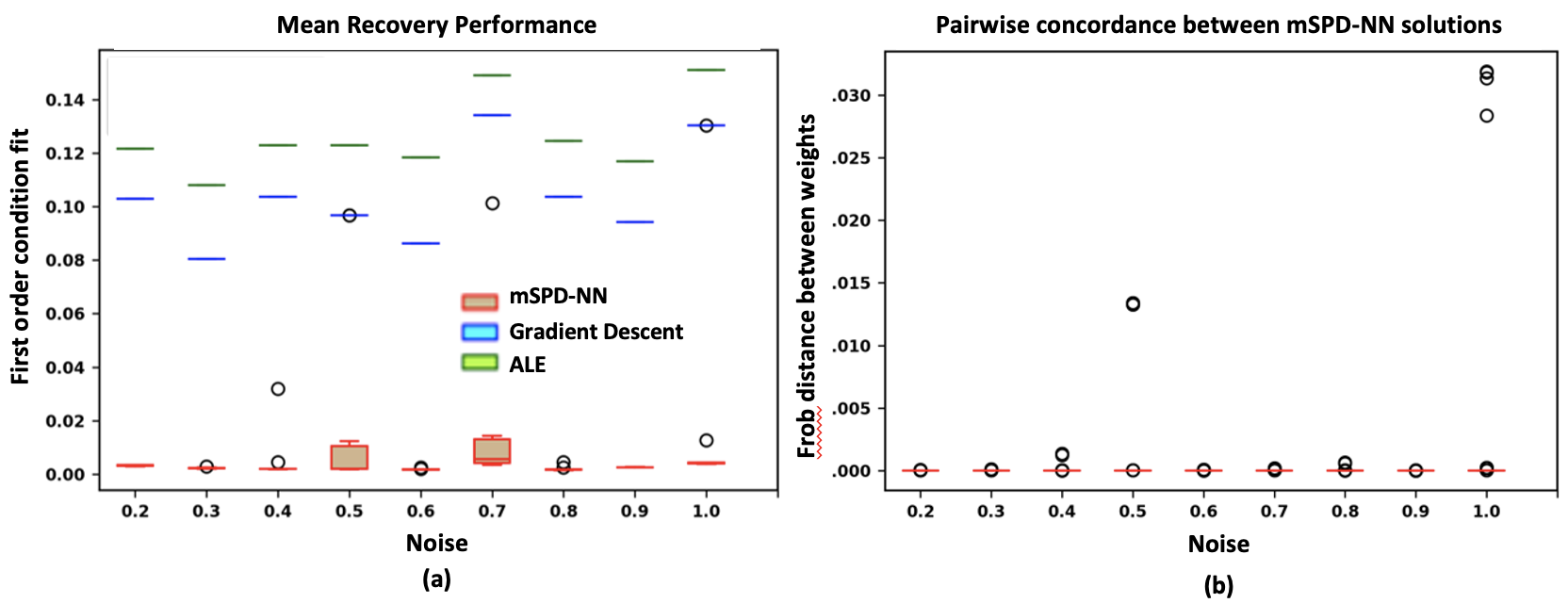}}
    \caption{Performance of the mSPD-NN, gradient descent and ALE estimation under increasing additive noise: \textbf{(a)} First order condition fit (Eq.~\ref{loss}) \textbf{(b)} Pairwise distance between the recovered mSPD-NN solutions across random initializations.}
    \label{fig:Geo-noise}
\end{figure}

Fig.~\ref{fig:Geo-noise}(a) illustrates the first-order condition fit $\mathcal{L}(\mathbf{G}^{\text{est}}_{R})$ across all three methods for increasing noise~$\sigma$. As seen, $\mathcal{L}(\mathbf{G}^{est}_{R})$ for the mSPD-NN is consistently lower than the corresponding value for the gradient descent and ALE algorithm, suggesting improved performance despite increasing corruption to the CPC process. The ALE algorithm is designed to utilize the CPC structure within the generating process, but its poor performance suggests that it is particularly susceptible to noise. Fig.~\ref{fig:Geo-noise}(b) plots the pairwise distances between the geodesic means estimated by mSPD-NN across the 10 random initializations. As seen, mSPD-NN produces a consistent solution, thus underscoring its robustness.

\subsection{Experiments on Functional Connectomics Data}
\label{FCCon}

\noindent\textbf{Dataset:} To probe the efficacy of the mSPD-NN for representation learning on real world matrix manifold data, we experiment on several groupwise discrimination tasks (such as group-wise discrimination, classification and clustering) on the publicly available CNI 2019 Challenge dataset~\cite{schirmer2021neuropsychiatric} consisting of preprocessed rs-fMRI time series, provided for $158$ subjects diagnosed with Attention Deficit Hyperactivity Disorder (ADHD), $92$ subjects with Autism Spectrum Disorder (ASD) with an ADHD comorbidity~\cite{leitner2014co}, and $257$ healthy controls. The scans were acquired on a Phillips 3T Achieva scanner using a single shot, partially parallel, gradient-recalled EPI sequence with TR/TE = $2500/30$ms, flip angle 70, voxel resolution = $3.05 \times 3.15 \times 3$mm, with a scan duration of either $128$ or $156$ time samples (TR). A detailed description of the demographics and preprocessing can be found in~\cite{schirmer2021neuropsychiatric}.
Connectomes are estimated via the Pearson's correlation matrix, regularized to be full-rank via two parcellations, the Automated Anatomical Atlas (AAL) ($P=116$) and the Craddocks~200 atlas ($P=200$).

\medskip
\noindent\textbf{Groupwise Discrimination:} We expect that FC differences between the ASD and ADHD cohorts are harder to tease apart than differences between ADHD and controls~\cite{schirmer2021neuropsychiatric,leitner2014co}.
We test this hypothesis by comparing the geodesic means estimated via mSPD-NN for the three cohorts. 
For robustness, we perform bootstrapped trials for mean estimation by sampling $25$ random subjects from a given group (ADHD/ASD/Controls).
We then compute the Riemannian distance $d(\mathbf{G}_{R}(\{\mathbf{\Gamma}_{g1}\}),\mathbf{G}_{R}(\{\mathbf{\Gamma}_{g2}\}))$ between the mSPD-NN means associated with groups~$g1$ and $g2$. A higher value of $d(\cdot,\cdot)$ implies a better separation between the groups. We also run a Wilcoxon signed rank test on the distribution of $d(\cdot,\cdot)$.  

\begin{figure}[t!]
    \centering
\includegraphics[width=0.8\textwidth]{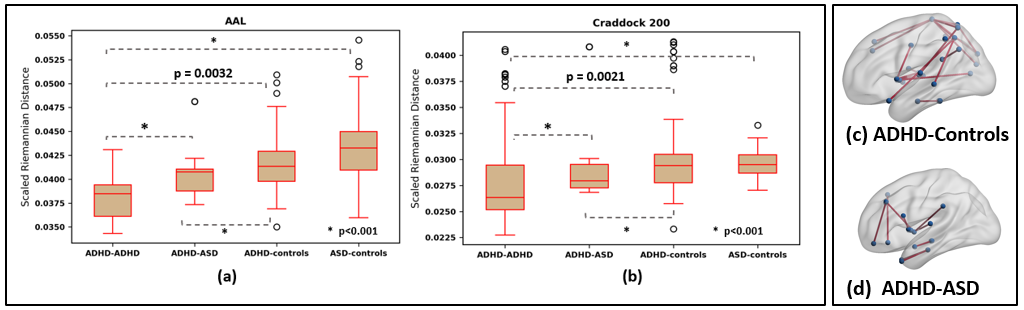}
    \caption{Groupwise discrimination between the FC matrices estimated via the \textbf{(a)} AAL \textbf{(b)} Craddock's 200 atlas, for the ADHD/ASD/Controls cohorts according to pairwise distances between the mSPD-NN mean estimates. Results of pairwise connectivity comparisons between group means for \textbf{(c)} ADHD-Controls \textbf{(d)} ADHD-ASD groups for the AAL parcellation. The red connections are significant differences ($p<0.001$).}
    \label{fig:Geo-FC}
\end{figure}

\begin{figure}[t!]
    \centering
    \includegraphics[width=0.75\textwidth]{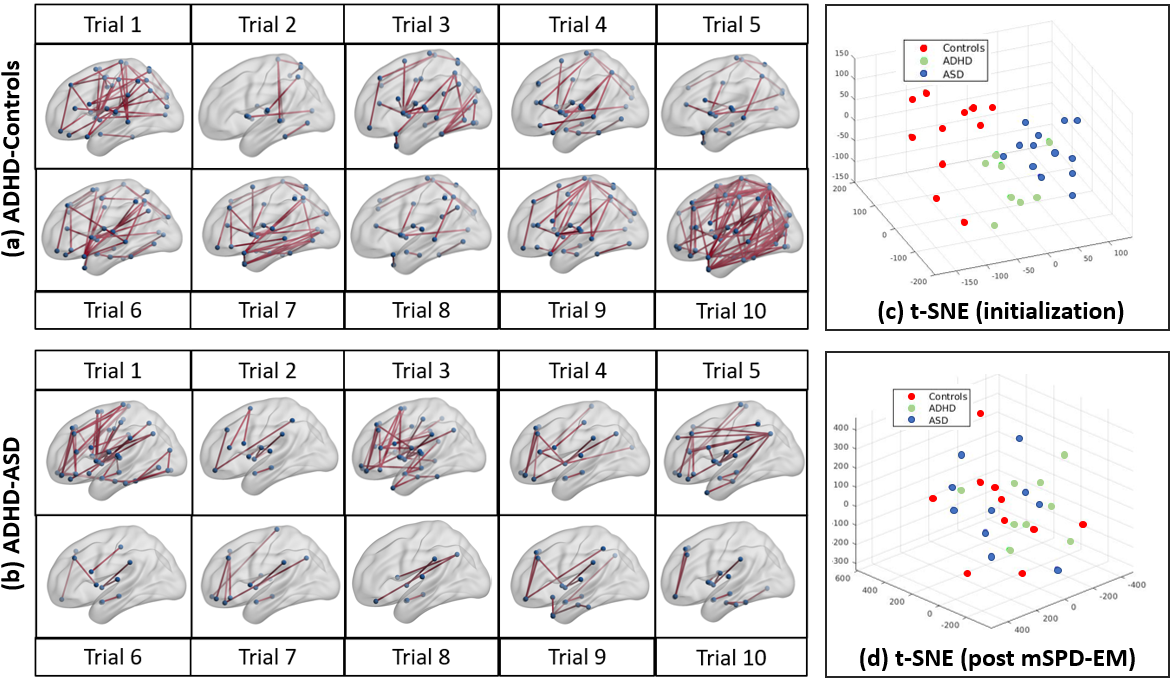}
    \caption{Pairwise differences between mSPD-NN group means for (a) ADHD-Controls (b) ADHD-ASD groups across bootstrapped trials. Significant differences marked in red ($p<0.001$). t-SNE plots for group means from experiment on (c) Groupwise Discrimination using mSPD-NN (d) After data-driven clustering via the mSPD-EM  }
    \label{fig:ADHD-Cont}
\end{figure}

Fig.~\ref{fig:Geo-FC} illustrates the pairwise distances between the geodesic means of cohorts $g1-g2$ across bootstrapped trials (t-SNE representations for the group means are provided in Fig.~\ref{fig:ADHD-Cont}(c)). As a sanity check, we note that the mean estimates across samples within the same cohort (ADHD-ADHD) are closer than those across cohorts (ADHD-controls, ASD-controls, ADHD-ASD). More interestingly, we observe that ADHD-controls separation is consistently larger than that of the ADHD-ASD groups for both parcellations. This result confirms the hypothesis that the overlapping diagnosis for the two classes translates to a reduced separability in the space of FC matrices and indicates that mSPD-NN is able to robustly uncover population level differences in FC.

\medskip
\noindent\textbf{Classification:} Building on the observation that mSPD-NN provides reliable group-separability, we adopt this framework for classification. Using the AAL parcellation, we randomly sample $25$ subjects from each class for training, and set aside the rest for evaluation with a $10\%/90\%$ validation/test split. We estimate the geodesic mean for each group across the training samples via $10$ bootstrapped trials, in which we sub-sample $80\%$ of the training subjects from the respective group. Permutation testing is performed on the mean estimates~\cite{zalesky2010network}, and functional connections (i.e., entries of $\mathbf{G}_{R}(\{\mathbf{\Gamma}_{n}\})$) that differ with an FDR-corrected threshold of $p<0.001$ are retained for classification. Finally, a Random Forest classifier is trained on the selected features to classify ADHD vs Controls. The train-validation-test splits are repeated $10$ times to compute confidence intervals. 

We use classification accuracy and area under the receiver operating curve (AU-ROC) as metrics for evaluation. The mSPD-NN feature selection plus Random Forest approach provides an accuracy of $0.62\pm{0.031}$ and an AU-ROC of $0.60\pm{0.04}$ for ADHD-Control classification on the test samples. We note that this approach outperforms all but one method on the CNI challenge leader-board~\cite{schirmer2021neuropsychiatric}. Moreover, one focus of the challenge is to observe how models trained on the ADHD vs Control discrimination task translate to ASD (with ADHD comorbidity) vs Control discrimination in a transfer learning setup. Accordingly, we apply the learned classifiers in each split to ASD vs Control classification and obtain an accuracy of $0.54\pm{0.044}$ and an AU-ROC of $0.53\pm{0.03}$. This result is on par with the best performing algorithm in the CNI-TL challenge. The drop in accuracy and AU-ROC for the transfer learning task is consistent with the performance profile of all the challenge submissions. These results suggest that despite the comorbidity, connectivity differences between the cohorts are subtle and hard to reliably capture. Nonetheless, the mSPD-NN+RF framework is a first step to underscoring stable, yet interpretable (see below) connectivity patterns that can discriminate between diseased and healthy populations. 

\medskip
\noindent\textbf{Qualitative Analysis:} To better understand the group-level connectivity differences, we plot the most consistently selected features (top 10 percent) from the previous experiment (ADHD-control feature selection) in Fig.~\ref{fig:Geo-FC}(c). We utilize the BrainNetViewer Software for visualization. The blue circles are the AAL nodes, while the solid lines denote edges between nodes. We observe that the highlighted connections appear to cluster in the sensorimotor and visual areas of the brain, along with a few temporal lobe contributions. Altered sensorimotor and visual functioning has been previously reported among children and young adults diagnosed with ADHD~\cite{duerden2012altered}. Adopting a similar procedure, we additionally highlight differences among the ASD and ADHD cohorts in Fig.~\ref{fig:Geo-FC}(d). The selected connections concentrate around the pre-frontal areas of the brain, which is believed to be associated with altered social-emotional regulation in Autism~\cite{pouw2013link}. We additionally provide an extended version of the group connectivity difference results across trials in Fig.~\ref{fig:ADHD-Cont} (a) ADHD vs Controls and (b) ADHD vs ASD. Across train-test-validation splits, we observe that several connectivity differences appear fairly consistently. Overall, the patterns highlighted via statistical comparisons on the mSPD-NN estimates are both robust as well as in line with the physiopathology of ADHD and ASD reported in the literature.

\medskip
\noindent\textbf{Data-Driven Clustering:} Finally, we evaluate the stability of the mapping between the functional connectivity and diagnostic spaces via a geometric clustering experiment. We use the geodesic mean estimates from the groupwise discrimination experiment (generated using the ground truth Controls/ASD/ADHD labels and mSPD-NN) as an initialization and track the shift in the diagnostic assignments upon running an unsupervised \textbf{E}xpectation-\textbf{M}aximization (EM) algorithm. At each iteration of the mSPD-EM, the E-Step assigns cluster memberships to a given subject according to the geodesic distance (Eq.~(\ref{dist})) from the cluster centroids, while the M-Step uses the mSPD-NN for recomputing the centroids. Upon  convergence, we evaluate the alignment between the inferred clusters and diagnostic labels. To this end, we map each cluster to a diagnostic label according to majority voting, and measure the cluster purity (fraction of cluster members that are correctly assigned). mSPD-EM provides an overall cluster purity of $0.59 \pm 0.05$ (Controls),  $0.52 \pm 0.12$ (ADHD), ASD $0.51 \pm 0.09$ (ASD), indicating that there is considerable shift in the assignment of diagnostic labels from ground truth. We also visualise the cluster centroids using t-Stochastic Neighbor Embeddings (t-SNE) at initialization and after convergence of the mSPD-EM in Fig.~\ref{fig:ADHD-Cont} (c) and (d) respectively. We provide 3-D plots to better visualise the cluster separation. Again, we observe that the diagnostic groups overlap considerably and are challenging to separate in the functional connectivity space alone. One possible explanation may be that the distinct neural phenotypes between the disorders are being overwhelemed by other rs-fMRI signatures. Given the migration of diagnostic assignments from the ground truth, the strict diagnostic criteria used to separate the diseased and healthy cohorts group may need to be more critically examined.

\section{Conclusion}
We have proposed a novel mSPD-NN framework to reliably estimate the geodesic mean of a collection of functional connectivity matrices. Through extensive simulation studies, we demonstrate that the mSPD-NN scales well to high-dimensional data and can handle input noise when compared with current iterative methods. By conducting a series of experiments on group-wise discrimination, feature selection, classification, and clustering, we demonstrate that the mSPD-NN is a reliable framework for discovering consistent group differences between patients diagnosed with ADHD-Autism comorbidities and controls. The mSPD-NN makes minimal assumptions about the data and can potentially be a useful tool to advance data-scientific and clinical research.

\paragraph{\textbf{Acknowledgements}} This work is supported by the National Science Foundation CAREER award 1845430 (PI Venkataraman), the National Institute of Health R01HD108790 (PI Venkataraman) and R01EB029977 (PI Caffo).
\bibliographystyle{splncs04}
\footnotesize{\bibliography{iclr2023_conference}}

\begin{thebibliography}{10}
\providecommand{\url}[1]{\texttt{#1}}
\providecommand{\urlprefix}{URL }
\providecommand{\doi}[1]{https://doi.org/#1}

\bibitem{banerjee2015nonlinear}
Banerjee, M., et~al.: Nonlinear regression on riemannian manifolds and its
  applications to neuro-image analysis. In: International Conference on Medical
  Image Computing and Computer-Assisted Intervention. pp. 719--727. Springer
  (2015)

\bibitem{bessadok2022graph}
Bessadok, A., Mahjoub, M.A., Rekik, I.: Graph neural networks in network
  neuroscience. IEEE Transactions on Pattern Analysis and Machine Intelligence
  (2022)

\bibitem{congedo2015approximate}
Congedo, M., Afsari, B., Barachant, A., Moakher, M.: Approximate joint
  diagonalization and geometric mean of symmetric positive definite matrices.
  PloS one  \textbf{10}(4),  e0121423 (2015)

\bibitem{congedo2017fixed}
Congedo, M., Barachant, A., Koopaei, E.K.: Fixed point algorithms for
  estimating power means of positive definite matrices. IEEE Transactions on
  Signal Processing  \textbf{65}(9),  2211--2220 (2017)

\bibitem{dong2017deep}
Dong, Z., et~al.: Deep manifold learning of symmetric positive definite
  matrices with application to face recognition. In: Thirty-First AAAI
  Conference on Artificial Intelligence (2017)

\bibitem{duerden2012altered}
Duerden, E.G., Tannock, R., Dockstader, C.: Altered cortical morphology in
  sensorimotor processing regions in adolescents and adults with
  attention-deficit/hyperactivity disorder. Brain research  \textbf{1445},
  82--91 (2012)

\bibitem{d2019integrating}
D’Souza, N.S., et~al.: Integrating neural networks and dictionary learning
  for multidimensional clinical characterizations from functional connectomics
  data. In: International Conference on Medical Image Computing and
  Computer-Assisted Intervention. pp. 709--717. Springer (2019)

\bibitem{d2020joint}
D’Souza, N.S., et~al.: A joint network optimization framework to predict
  clinical severity from resting state functional mri data. NeuroImage
  \textbf{206},  116314 (2020)

\bibitem{d2021matrix}
D’Souza, N.S., et~al.: A matrix autoencoder framework to align the functional
  and structural connectivity manifolds as guided by behavioral phenotypes. In:
  International Conference on Medical Image Computing and Computer-Assisted
  Intervention. pp. 625--636. Springer (2021)

\bibitem{fornito2013graph}
Fornito, A., Zalesky, A., Breakspear, M.: Graph analysis of the human
  connectome: promise, progress, and pitfalls. Neuroimage  \textbf{80},
  426--444 (2013)

\bibitem{fox2007spontaneous}
Fox, M.D., et~al.: Spontaneous fluctuations in brain activity observed with
  functional magnetic resonance imaging. Nat. Rev. Neuro.  \textbf{8}(9), ~700
  (2007)

\bibitem{jeuris2015riemannian}
Jeuris, B.: Riemannian optimization for averaging positive definite matrices
  (2015)

\bibitem{jolliffe2016principal}
Jolliffe, I.T., Cadima, J.: Principal component analysis: a review and recent
  developments. Philosophical Transactions of the Royal Society A:
  Mathematical, Physical and Engineering Sciences  \textbf{374}(2065),
  20150202 (2016)

\bibitem{khosla2019machine}
Khosla, M., et~al.: Machine learning in resting-state fmri analysis. Magnetic
  resonance imaging  \textbf{64},  101--121 (2019)

\bibitem{leitner2014co}
Leitner, Y.: The co-occurrence of autism and attention deficit hyperactivity
  disorder in children--what do we know? Frontiers in human neuroscience
  \textbf{8}, ~268 (2014)

\bibitem{lindquist2008statistical}
Lindquist, M.A.: The statistical analysis of fmri data. Statistical science
  \textbf{23}(4),  439--464 (2008)

\bibitem{moakher2005differential}
Moakher, M.: A differential geometric approach to the geometric mean of
  symmetric positive-definite matrices. SIAM journal on matrix analysis and
  applications  \textbf{26}(3),  735--747 (2005)

\bibitem{nandakumar2020multi}
Nandakumar, N., et~al.: A multi-task deep learning framework to localize the
  eloquent cortex in brain tumor patients using dynamic functional
  connectivity. In: MICCAI Workshop on Machine Learning in Clinical
  Neuroimaging, pp. 34--44. Springer (2020)

\bibitem{nguyen2019neural}
Nguyen, X.S., et~al.: A neural network based on spd manifold learning for
  skeleton-based hand gesture recognition. In: Proceedings of the IEEE/CVF
  Conference on Computer Vision and Pattern Recognition. pp. 12036--12045
  (2019)

\bibitem{pennec2006riemannian}
Pennec, X., Fillard, P., Ayache, N.: A riemannian framework for tensor
  computing. International Journal of computer vision  \textbf{66}(1),  41--66
  (2006)

\bibitem{pham2001joint}
Pham, D.T.: Joint approximate diagonalization of positive definite hermitian
  matrices. SIAM Journal on Matrix Analysis and Applications  \textbf{22}(4),
  1136--1152 (2001)

\bibitem{pouw2013link}
Pouw, L.B., et~al.: The link between emotion regulation, social functioning,
  and depression in boys with asd. Research in Autism Spectrum Disorders
  \textbf{7}(4),  549--556 (2013)

\bibitem{schirmer2021neuropsychiatric}
Schirmer, M.D., et~al.: Neuropsychiatric disease classification using
  functional connectomics-results of the connectomics in neuroimaging transfer
  learning challenge. Medical image analysis  \textbf{70},  101972 (2021)

\bibitem{zalesky2010network}
Zalesky, A., Fornito, A., Bullmore, E.T.: Network-based statistic: identifying
  differences in brain networks. Neuroimage  \textbf{53}(4),  1197--1207 (2010)

\bibitem{zhang2013majorization}
Zhang, T.: A majorization-minimization algorithm for the karcher mean of
  positive definite matrices. arXiv preprint arXiv:1312.4654  (2013)

\end{thebibliography}
\end{document}